\documentclass[10pt,twocolumn,superscriptaddress,preprintnumbers,nofootinbib]{revtex4-1}
\usepackage[T1]{fontenc} 
\usepackage{amssymb}
\usepackage{amsmath}
\usepackage{color}
\usepackage{xspace}
\usepackage{appendix}
\usepackage{float}
\usepackage{fancyhdr}
\usepackage{hyperref}
\usepackage{multirow}
\usepackage{diagbox}
\usepackage[utf8x]{inputenc}
\usepackage{soul}
\usepackage[compat=1.1.0]{tikz-feynhand}
\usepackage{subfig}
\usepackage{kantlipsum}

\begin{document}


\title{ Mass suppression effect in QCD radiation and hadron angular distribution in jet}
\author{Chuan-Hui Jiang}
\email{jiangch@mail.sdu.edu.cn}
\affiliation{School of Physics, Shandong University, Jinan, Shandong 250100, China}
\author{Hai Tao Li}
\email{haitao.li@sdu.edu.cn}
\affiliation{School of Physics, Shandong University, Jinan, Shandong 250100, China}
\author{Shi-Yuan Li}
\email{lishy@sdu.edu.cn}
\affiliation{School of Physics, Shandong University, Jinan, Shandong 250100, China}
\author{Zong-Guo Si}
\email{zgsi@sdu.edu.cn}
\affiliation{School of Physics, Shandong University, Jinan, Shandong 250100, China}

\begin{abstract}
The finite mass of the heavy quark suppresses the collimated radiations; this is generally referred to as
the dead cone effect. In this paper, we study the distribution of hadron multiplicity over the hadron opening angle
with respect to the jet axis for various jet flavors. The corresponding measurement can be the most straightforward
and simplest approach to explore the dynamical evolution of the radiations in the corresponding jet, which can expose
the mass effect. We also propose a transverse energy-weighted angular distribution, which sheds light on the interplay
between perturbative and non-perturbative effects in the radiation. Through Monte-Carlo simulations, our
calculations show that the dead cone effect can be clearly observed by finding the ratio between the b and light-quark
(inclusive) jets; this is expected to be measured at the LHC in the future.
\end{abstract}

\maketitle


\section{Introduction}

Measuring jet production offers unique opportunities to study the perturbative and non-perturbative behavior of QCD~\cite{Larkoski:2017jix,Kogler:2018hem,Marzani:2019hun}. Heavy flavor jets have been used to verify
perturbative QCD and explore the non-perturbative effect,
thereby enhancing our understanding of jet evolution
in vacuum or QCD media~\cite{Cacciari:1998it,Norrbin:2000zc}. Currently at the
LHC, identification techniques of heavy flavor jets make
it possible to discriminate the jets originating from $b$ or $c$
quarks and those from light flavor quarks or gluons. Experimental
collaborations at the LHC have measured
heavy flavor jet production~\cite{CMS:2012pgw,CMS:2013qak,CMS:2015gcq, CMS:2016wma,CMS:2017wtu,CMS:2018dqf,CMS:2019jis,ALICE:2021aqk,ALICE:2022mur,ALICE:2022phr,ATLAS:2022fgb} and more results are
expected in the near future, especially from LHC Run 3.

Many efforts on the theoretical side have been devoted
to studying heavy flavor jets at various colliders.
The inclusive $p_T$ spectrum of heavy flavor jets~\cite{Dai:2018ywt,Li:2018xuv} can be predicted with the help of the semi-inclusive jet function~\cite{Kang:2016mcy,Dai:2016hzf}. The heavy flavor jet with high transverse
momentum can be used to understand the evolution
of a massless quark when the mass is small enough to be
ignored in comparison to the jet energy. In addition, the
mass impact on the perturbative and non-perturbative
nature during the evolution of heavy flavor quarks can be
monitored by measuring the radiation pattern inside jets.
In this regard, the most famous phenomenon is the dead
cone effect~\cite{Dokshitzer:1991fc,Dokshitzer:1991fd,Schumm:1992xt}, which is a direct consequence of the
suppression of the collinear radiation due to the mass of
the {\it radiator}, i.e., the heavy quark. In recent years, the
mass effect using heavy flavor jets has drawn a lot of attention
in both theoretical and experimental studies. This
mass effect has been analyzed in gauge theory models
such as QED and QCD~\cite{Maltoni:2016ays,Craft:2022kdo,Jiang:2023pvj,Ghira:2023bxr,Caletti:2023spr,Ba:2023hix,Kluth:2023rst}.  Besides, many studies
have been devoted to quantifying the dead cone effect in
heavy-ion and electron-ion collisions~\cite{Li:2018xuv,Wang:2020ukj,Dai:2021mxb,Dang:2023tmb,Cunqueiro:2022svx,Prakash:2023wbs,Wang:2023eer,Andres:2023ymw}.

In general, the non-zero mass of heavy quarks can
control the infrared behavior of the radiation, leading to a
specific perturbative radiation effect. In the collinear limit,
the splitting of a massive quark can be described by effective
field theory as~\cite{Kang:2016ofv} 
\begin{align}
    \left[\frac{dN}{d^2\mathbf{k}_\perp d z }\right]_{Q \to Qg} \propto  \frac{1}{\mathbf{k}^2_\perp + z^2 m^2}\;,
    \label{eq:1}
\end{align}
where $\mathbf{k}_\perp$ is the transverse momentum of the emitted gluon and $z$ is the energy fraction of the gluon relative to the parent massive quark. According to Eq.~(\ref{eq:1}) if we compare with the  massless quark splitting, the small-angle radiation is significantly suppressed by the mass term, {\it i.e.}, the dead cone effect \cite{Dokshitzer:1991fd}. 
Jet properties and  substructures have been widely used to study this effect. In~\cite{Chang:2017gkt,Li:2017wwc,Caletti:2023spr} the authors investigated  $z_g$ and $\theta_g$ distributions for the groomed b jet that the b-quark mass effect plays an important role in the collinear splitting pattern.  Many other developments have been proposed on jet substructures for heavy flavor jet production~\cite{Makris:2018npl,Lee:2019lge,Cunqueiro:2022svx}.  There are also studies that applied heavy
flavor jet observables at the electron-ion collider~\cite{Li:2021gjw}; see also Ref.~\cite{AbdulKhalek:2022hcn} and the references therein. By inspecting
the splitting inside the jet, the dead cone effect has
already been measured using the substructure of the
charm jet by ALICE~\cite{ALICE:2021aqk,ALICE:2022mur,ALICE:2022phr}. 

Eq.~(\ref{eq:1}) also shows that generally the mass effect is
more significant at relatively low energy scales/small
angles. To obtain a better energy/angular resolution for
this effect, the most straightforward approach is to measure
the angular distribution of hadron multiplicity in the
jet. The dead cone effect can be explored by comparing
the hadron angular distribution between heavy flavors
and light quark jets. As a physical phenomenon that typically
occurs at preconfinement scale  \cite{Amati:1979fg,Dokshitzer:1991wu}, hadron
multiplicity is not an infrared safe observable. However,
we expect that the angular distribution of the hadrons preserves
most of the perturbative effect from the heavy
quark mass. We also propose a transverse energy weighted
angular distribution that connects the perturbative
and non-perturbative multiplicity distributions. The
proposed observable can be measured at the LHC and
provides a new approach for investigating the mass effect
for QCD radiation and jet formation. Furthermore,
this observable can be used to test or tune the hadronization
models of Monte-Carlo event generators.

The rest of the paper is organized as follows. We
provide a definition of the observable in Sec.~\ref{sec:obs}. Section ~\ref{sec:nums} presents numerical results from Monte-Carlo event
generators and discusses the mass effect on jet evolution. Section~\ref{sec:concl} concludes the paper..

\section{Definition of the Observable} \label{sec:obs}

In general, the non-zero mass of a radiator or the radiated particles can control the infrared behavior of the radiation. Given that heavy quark mass does not originate from confinement dynamics, it can lead to a specific perturbative radiation effect.  The averaged charged particle multiplicities in $e^+e^-$ collision were used to investigate the mass effect~\cite{Kiselev:1988nm, Schumm:1992xt,Petrov:1994ng, Dokshitzer:2005ri}. Currently, leveraging the LHC and high luminosity LHC, it is possible to take a closer look inside jets. In this paper, we propose the simplest approach to comprehensively analyze the dead cone effect for heavy flavor jets: an averaged multiplicity distribution defined as 
\begin{align}
    \frac{d\langle N_{ch} \rangle }{d\theta} =  \sum_{\rm ch \in jet}  \frac{dP_{ch}}{d\theta}\;,
    \label{eq:def}
\end{align}
where $\theta$ is the opening angle between the jet axis and the moving direction of the charged hadron $ch$ and $dP_{ch}/d\theta$ is the probability distribution as a function of $\theta$ for a charged hadron $ch$.  According to the dead cone effect the collinear radiation inside heavy flavor jets is suppressed at a small angle. A detailed study of $d\langle N_{ch} \rangle/d\theta$ for jets with various transverse momenta can be used to quantitatively identify the energy scale of the dead cone effect. Although this observable is not infrared safe, the distribution is expected to reflect the pattern from perturbative radiations; in particular, the difference between heavy flavor and light quark jets. This observable may depend on the definition of the jet axis. In this study, we used the traditional energy combination scheme to retain the correlation between the directions of the jet axis and momentum of the parent parton.

Regarding the non-perturbative nature of $d\langle N_{ch} \rangle/d\theta$ , it is interesting to investigate to what extent this angle distribution can be affected by nonperturbative QCD. Therefore, we introduce a variation to Eq.~(\ref{eq:def}) 
\begin{align}
     \frac{d\langle N_{ch} \rangle (\kappa) }{d\theta}=  \sum_{\rm ch \in jet} \left(\frac{p_{T,{\rm ch}}}{p_{T,{\rm jet}}}\right)^\kappa \frac{dP_{ch}}{d\theta} \;,
    \label{eq:defk}
\end{align}
where $p_{T,ch}$ and $p_{T,j}$ are the transverse momenta for the charged hadron and jets, respectively; $\kappa$ is a free parameter. When $\kappa=0$, it is reduced to the multiplicity distributions inside jets. For  $\kappa=1$, $d\langle N_{ch}\rangle(\kappa=1)/d\theta$ measures the $\theta$ dependence of the energy deposit inside the jet cone and  $\int_0^{r} d\theta \frac{d\langle N_{ch}\rangle(\kappa=1)}{d\theta}$  corresponds to the infrared safe observable jet shape, or jet transverse energy profile~\cite{Ellis:1992qq}. For a variation in the range of $0\leq \kappa \leq 1$  a bridge between infrared unsafe and safe observables is established, which can be utilized to probe the nonpertubative effect.

\section{Numerical results}\label{sec:nums}

\begin{figure*}
  \centering
  \includegraphics[width=0.9\textwidth]{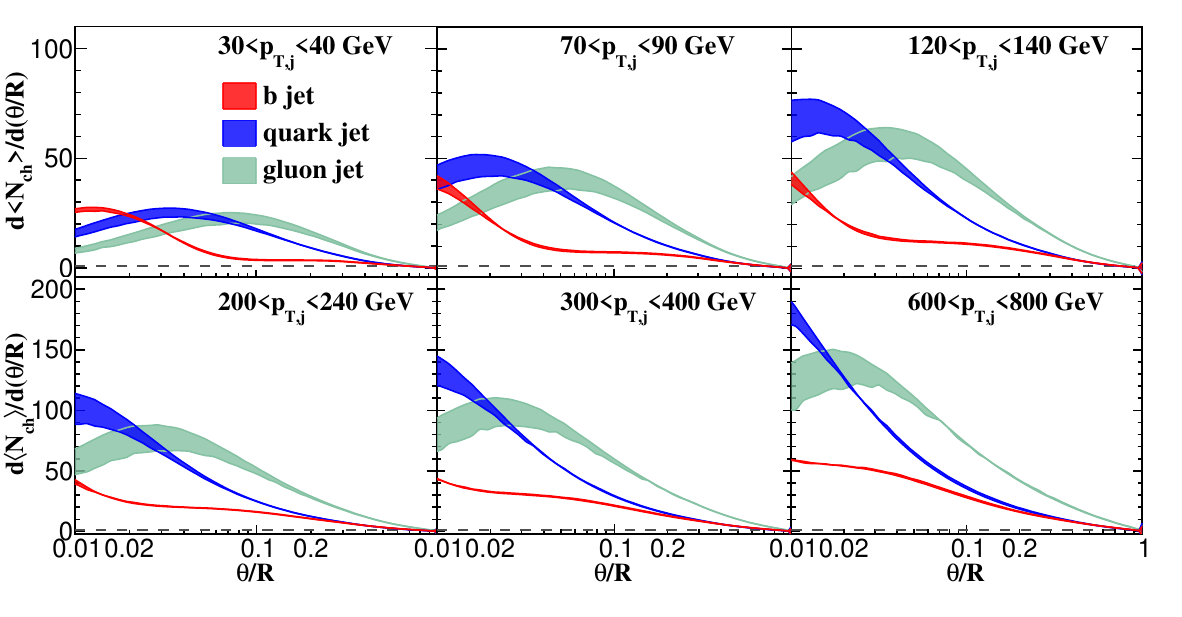}
  \caption{ Angle distribution of charged-particle multiplicity in different intervals of jet transverse momentum.  The x-axis is the opening angle between the charged particle and the jet axis which is normalized by the jet radius $R=0.4$. The red, blue, and green bands represent the b, quark, and gluon jets. The bands are Monte Carlo uncertainties. The black lines are the multiplicity distributions for b-jet without counting b hadrons.}
    \label{fig:n-distri}
\end{figure*}

Numerical results for $d\langle N_{ch}\rangle(\kappa)/d\theta$ are obtained from simulations with {\tt PYTHIA}~\cite{Sjostrand:2014zea,Bierlich:2022pfr}. The default setting in {\tt PYTHIA 8.306} was adopted for parton showers and hadronizaiton. The simulation was performed for dijet production at the 13 TeV LHC.   To investigate the physics of jets with different flavors, the multiparton interaction was switched off. The jets were constructed with anti-$k_T$ algorithms~\cite{Cacciari:2008gp}   and jet radius $R=0.4$ using only the charged tracks with $p_{T}>1$ GeV and $|\eta|<2.0$.  The recombination of jets was achieved using the package {\tt FastJet}~\cite{Cacciari:2011ma}. We classify the jet as a b jet if there was at least one B hadron in its component.

Figure~\ref{fig:n-distri} shows the distributions of $d\langle N_{ch}\rangle/d\theta$ for quark, gluon, and $b$ jets for various intervals of jet transverse momentum. The width of the bands corresponds to the uncertainties from varying the scales in parton showers by a factor of two. The integration over the $\theta/R$ axis provides the average charged multiplicity of the jet. As expected, he radiation from light quarks and gluons is greater.  In particular, there is more radiation for gluon jets at large angles, leading to a broader distribution. At a large angle, the $b$ jet behaves like the light quark jet; this is more evident for the high $p_T$ jet.  A clear suppression of the radiation can be observed for $b$ jet at small angle radiation. Needless to say, the suppression stems from the  dead cone effect.  Roughly speaking, the distributions for the quark and $b$ jet are expected to be different  when the typical scale of the collinear splitting inside jets $ p_{T,j} \theta\ z (1-z) \leq p_{T,j} \theta/4$\footnote{where $z$  is the momentum fraction of final state parton comparing to the parent parton} is close to the $b$ quark mass $m_b$.    In the first and last panels of Fig.~\ref{fig:n-distri} the black lines show the multiplicity distribution for b jets if we do not count the $b$ hadrons inside jets. We find that $b$ hadron tends to stay close to the jet axis.  As a result, there is a growing trend for very small angles,  particularly when the jet $p_{T,j}$ is small.

\begin{figure*}
  \centering
  \includegraphics[width=0.8\textwidth]{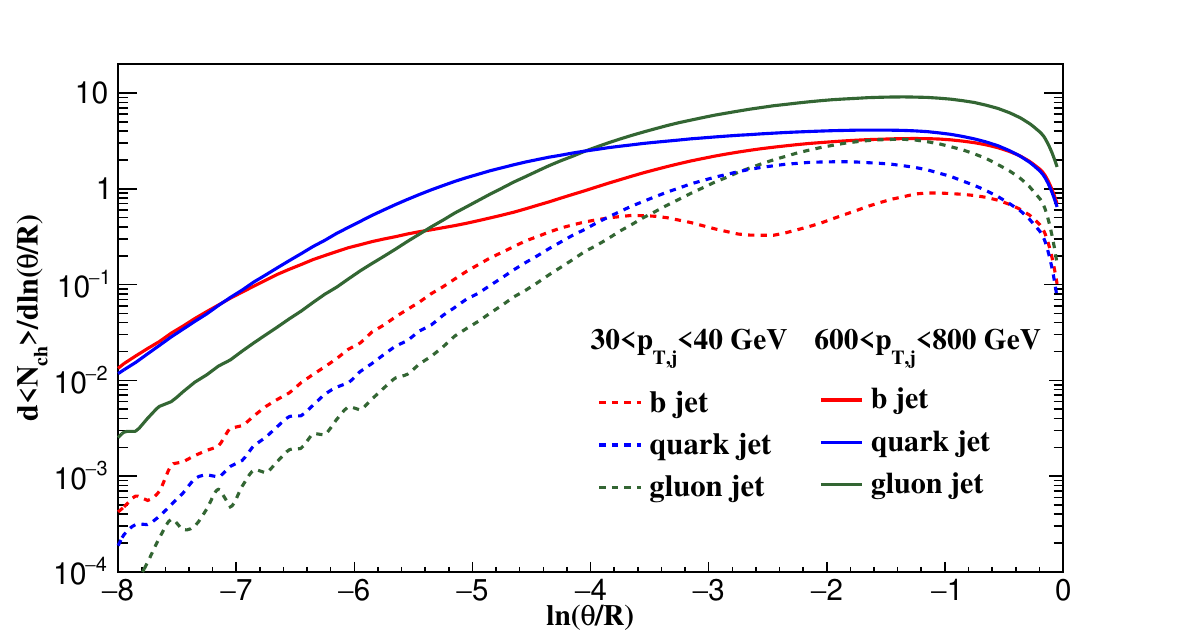}
   \caption{Multiplicity distribution over angles between changed particles and jet axis with jet transverse momenta $30<p_{T,j}<40$ GeV (dashed lines) and $600<p_{T,j}<800$ GeV  (solid lines).}
    \label{fig:lntheta}
\end{figure*}

\begin{figure*}
  \centering
  \includegraphics[width=0.49\textwidth]{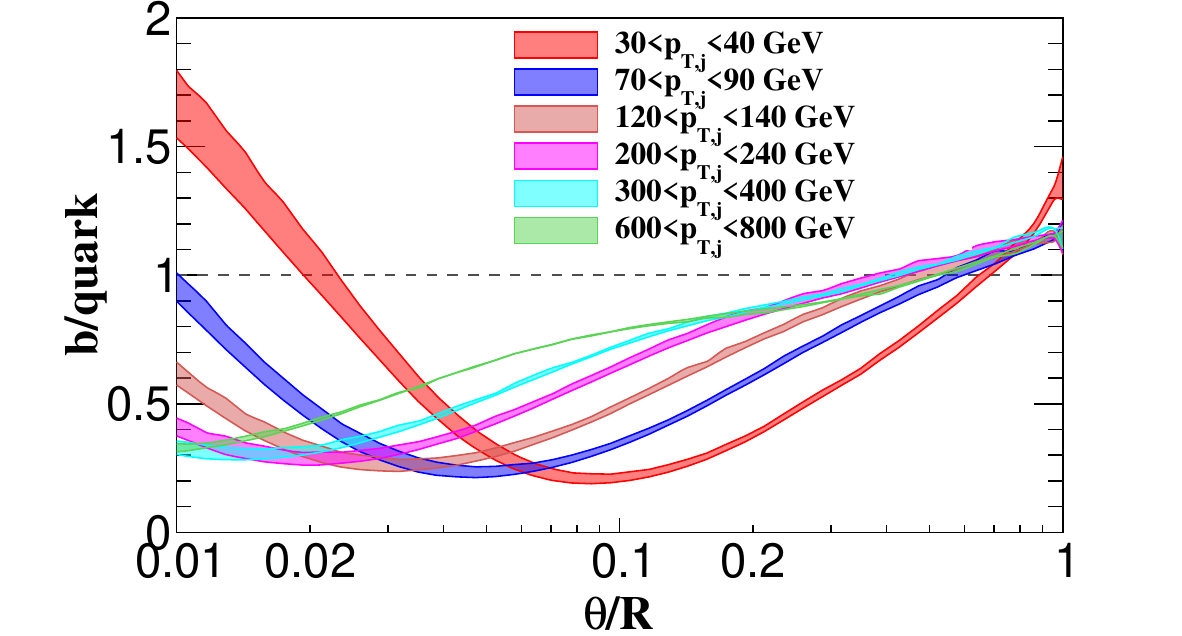}
  \includegraphics[width=0.49\textwidth]{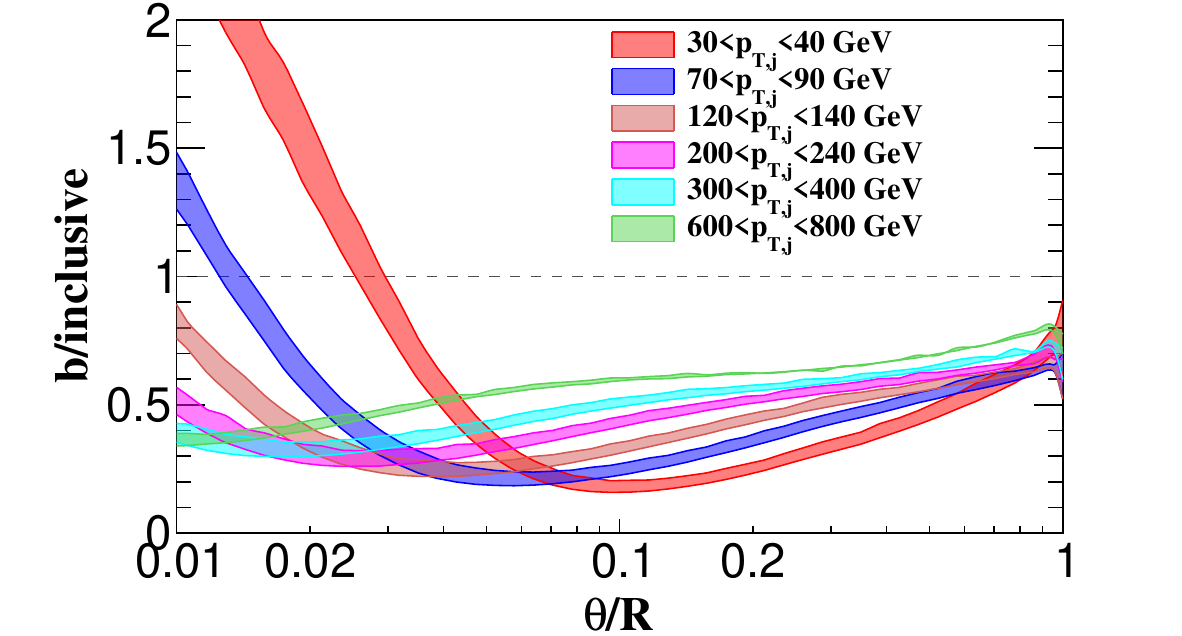}
\caption{ Ratios of the multiplicity distributions of b-jet to quark jet (left) and b jet to inclusive jet (right) with various jet transverse momenta $p_{T,j}$.}
    \label{fig:ratio}
\end{figure*}

\begin{figure*}
  \centering
  \includegraphics[width=0.49\textwidth]{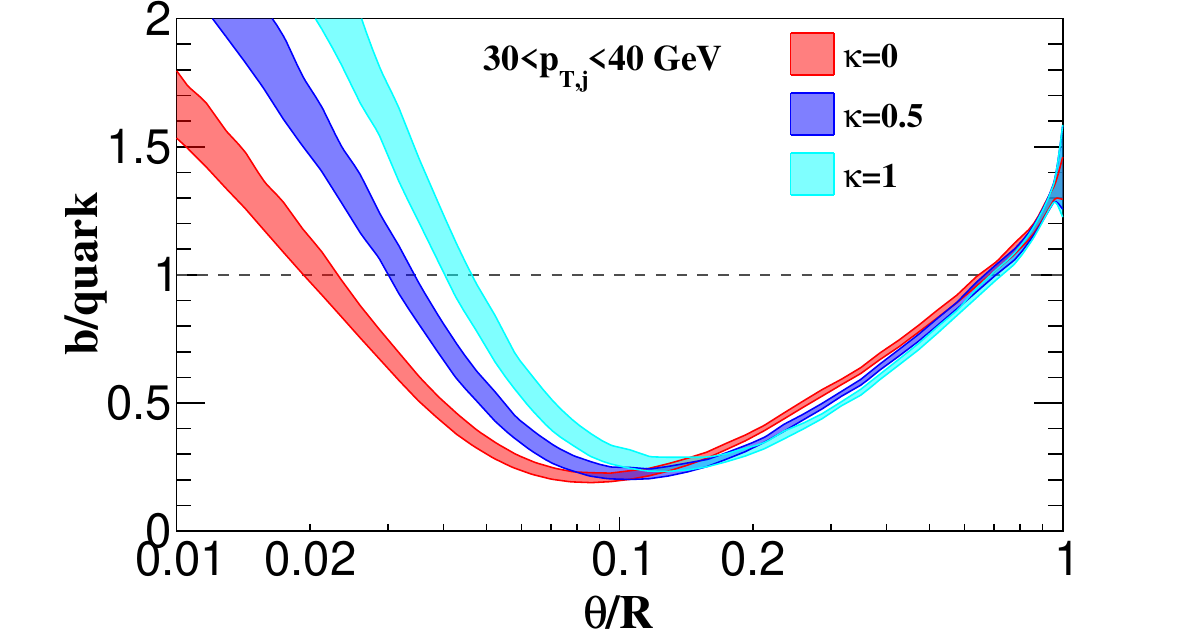}
  \includegraphics[width=0.49\textwidth]{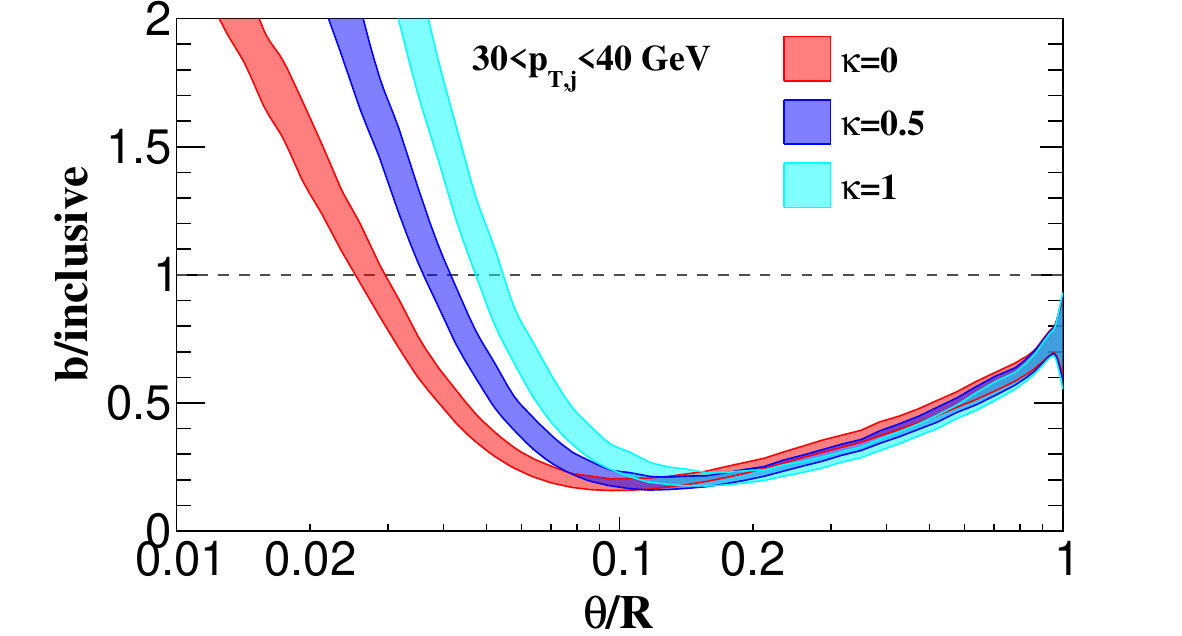}
\caption{The angle dependence of the charged-particle multiplicity ratio of b-jet to quark jet (left) and inclusive jet (right) with jet transverse momenta $30<p_{T,j}<40 \mathrm{GeV}$ and $|\eta_{jet}|<2.0$ varying $\kappa$ values.}
    \label{fig:kappa}
\end{figure*}

One of the interesting features of the multiplicity distribution is the scaling behavior in the limit $\theta\to0$. Figure~\ref{fig:lntheta} shows the multiplicity distributions for extremely small opening angles for jets with the transverse momentum  $30<p_{T,j} <40$ GeV and  $600<p_{T,j} <800$  GeV. When $p_{T,j} \theta \leq \Lambda_{\rm QCD}$, the phase space is extremely limited and the distributions are supposed to be dominated by non-perturbative features of QCD. Surprisingly, for jet multiplicities, we find that the quark, gluon, and b jets exhibit a similar scaling behavior in the small $\theta$ region.  This phenomenon can be explained by the existence of uniformly distributed hadrons in the collinear limit of jets;  as a result $d\langle N_{ch}\rangle/d\theta \propto \theta$, similar to energy-energy-correlators first reported by Ref.~\cite{Komiske:2022enw}. 
This interesting feature might originate from kinematics, given that the dynamic evolu- tion is frozen below $\Lambda_{\rm QCD}$.  
Consistent with what we find in Fig.~\ref{fig:n-distri}, the multiplicity distribution of the b jet is suppressed and then enhanced when decreasing the angle relative to the jet axis. The enhancement in the small angle for the $b$ jet mainly comes from that the B hadron tends to stay close to the jet axis. By varying $\kappa$ in Eq.~(\ref{eq:def}) we observe relatively small $\kappa$ dependence in the small angle region for $b$ jet. From Figs.~\ref{fig:n-distri} and ~\ref{fig:lntheta} it can be concluded that  the $b$-jet behaves like a $B$ hadron dressed with relatively soft radiations for lower $p_T$ jets.

To investigate the mass effect in detail, we present the ratio of the multiplicity distributions between $b$  and quark/inclusive jets at the LHC in Fig.~\ref{fig:ratio}.  
The mass effect is supposed to become smaller with increasing $\theta$ because the typical QCD scale for the splitting is larger, while at an extremely small angle, there is an enhancement observed for heavy flavor jet given that the collinear radiation is suppressed and the parent particle tends to stay close to the jet axis.  As a result of the interplay between these effects, there are dips in the distributions of the ratio. The dead cone effect can be captured by the position of the dip, which depends on the transverse momentum of the jet.  
The right plot of fig.~\ref{fig:ratio} displays the ratios between the distributions of $b$ and inclusive jets.   We found that the ratios are similar to those between the $b$  and quark jet; this is an approach to test the dead cone effects directly. A  measurement of the multiplicity distributions of jets can be used to investigate the mass effect on the QCD dynamic evolution. 

For $\kappa<1$,  $d\langle N_{ch}\rangle/d\theta$  is not an infrared-safe observable, while for $\kappa=1$ $d\langle N_{ch}\rangle/d\theta$ corresponds to the differential jet shape and perturbatively calculable.   The quantitative effect from non-perturbative physics can be explored by varying $\kappa$ in Eq.~(\ref{eq:def}). The distribution of $d\langle N_{ch}\rangle(\kappa)/d\theta$  heavily depends on $\kappa$; for larger values of $\kappa$, the distribution is smaller. To further analyze the b-quark mass and the nonperturbative effect,  we present the dependence of the ratio between b-jet and light-quark/inclusive jet with $30 < p_{T,j}< 40 \rm{GeV}$  in Fig.~\ref{fig:kappa}. Remarkably, we found that in the large $\theta$ region the nonperturbative effect seems to be canceled in the ratios,  thereby setting a guideline for the nonperturbative corrections in heavy flavor jet substructures. In the small $\theta$ region the differences between distinct $\kappa$ settings are large, as expected. The results for different values of $\kappa$ demonstrate a good and smooth transition of non-perturbative and perturbative QCD. As shown in the right plot of fig.~\ref{fig:kappa}, the ratio between $b$ jet and inclusive jets still keeps the mass effect as for the case of the ratio between $b$ and quark jets.

\section{Conclusion}
\label{sec:concl}

In this paper, we present the most straightforward and simplest approach to expose the dead cone effect of heavy flavor jets: a multiplicity distribution over the opening angle between the hadron inside the jet and the jet axis. Although this multiplicity is not infrared safe, we expect that this distribution reveals the mass effect on both perturbative and non-perturbative evolutions of the heavy flavor jets. To address the non-perturbative effect, we propose a transverse energy weighted multiplicity distribution that sheds light on the interplay between perturbative
and non-perturbative effects.

We present simulations of the multiplicity distributions
for $b$, light quark, and gluon originated jets with  {\tt PYTHIA} using charged particles in the events. In comparison to light quark and gluon jets, the radiation of b jets is suppressed for $\theta\approx m_b/p_{Tj}$; this is the dead-cone effect. Given the $b$ jets resemble a hard B hadron dressed
with some soft radiations, we observed an enhancement
of the distribution and a scaling behavior indicating uniformly
distributed hadrons at an extremely small angle.
By calculating the ratio between the $b$ jet and light quark
or inclusive jets, the dead cone effect can be clearly observed
in the simulations and is expected to be measured at the LHC in the future.  
We also examined the $\kappa$ dependence
of the ratio of the multiplicity distributions between
the b jet and light quark or inclusive jets, and found that
the non-perturbative effect cancels at large angles and is
important in small-angle regions. One important fact is
that the dead cone or mass suppression on small-angle radiation
is not removed by increasing the quark energy,
which is also clearly observed from the $b$ fragmentation
function (see recent study~\cite{Kluth:2023umf}).

Last but not least, we remark that the multiplicity distribution
reveals the mass effect during the dynamical
evolution of heavy flavor jets. The shape of the distributions
indicates the energy scale of the dead cone effect. It
would be interesting to apply this observable to heavy ion
collisions, which can be used to reveal the scale of the interactions
of colored partons with quark-gluon plasma.

\section*{Acknowledgments}
We would like to thank the members of ITP at Shandong University for the useful discussion.  This work was supported by the National Science Foundation of China under Grant Nos. 12235008, 12275156, 12275157,  and 12321005.

\bibliographystyle{elsarticle-num}
\bibliography{bib}
\end{document}